\documentclass[conference]{IEEEtran}

\usepackage{setspace}

\usepackage{graphics,
           psfrag,
           epsfig,
           amsthm,
           cite,
           amssymb,
           url,
           dsfont,
           subfigure,
           algorithm,
           algorithmic,
           balance,
           enumerate,
           color,
           setspace
}
\usepackage{amsmath}

\usepackage{epstopdf}

\newtheorem{definition}{Definition}

\newcommand{\eref}[1]{(\ref{#1})}
\newcommand{\sref}[1]{Section~\ref{#1}}

\newcommand{\fref}[1]{Figure~\ref{#1}}

\newcommand{\cref}[1]{Constraint~\ref{#1}}

\newcommand{\ignore}[1]{}

\IEEEoverridecommandlockouts
\begin{document}

\title{On Minimizing the Maximum Broadcast Decoding Delay for Instantly Decodable Network Coding}

\author{
   \authorblockN{Ahmed Douik$^{\dagger}$, Sameh Sorour$^\ast$, Mohamed-Slim Alouini$^\dagger$, and Tareq Y. Al-Naffouri$^{\dagger\ast}$\\}%
   \authorblockA{$^\dagger$King Abdullah University of Science and Technology (KAUST), Kingdom of Saudi Arabia \\
    $^\ast$King Fahd University of Petroleum and Minerals (KFUPM), Kingdom of Saudi Arabia \\
    Email: $^\dagger$\{ahmed.douik,slim.alouini,tareq.alnaffouri\}@kaust.edu.sa \\
    $^\ast$\{samehsorour,naffouri\}@kfupm.edu.sa}
    }

\maketitle

\IEEEoverridecommandlockouts

\begin{abstract}
In this paper, we consider the problem of minimizing the maximum broadcast decoding delay experienced by all the receivers of generalized instantly decodable network coding (IDNC). Unlike the sum decoding delay, the maximum decoding delay as a definition of delay for IDNC allows a more equitable distribution of the delays between the different receivers and thus a better Quality of Service (QoS). In order to solve this problem, we first derive the expressions for the probability distributions of maximum decoding delay increments. Given these expressions, we formulate the problem as a maximum weight clique problem in the IDNC graph. Although this problem is known to be NP-hard, we design a greedy algorithm to perform effective packet selection. Through extensive simulations, we compare the sum decoding delay and the max decoding delay experienced when applying the policies to minimize the sum decoding delay \cite{ref2} and our policy to reduce the max decoding delay. Simulations results show that our policy gives a good agreement among all the delay aspects in all situations and outperforms the sum decoding delay policy to effectively minimize the sum decoding delay when the channel conditions become harsher. They also show that our definition of delay significantly improve the number of served receivers when they are subject to strict delay constraints.
\end{abstract}


\section{Introduction} \label{sec:intro}

\emph{Network Coding (NC)} gained much attention in the past decade. NC is a young field of study which birth is generally attributed to the seminal paper \cite{850663}. It was shown to be a promising solution to substantially improve the network capacity or the throughput and the delay over broadcast erasure channels. These merits are of great interest for the spread and proliferation of real-time applications and streaming requiring quick and reliable packet transmission over lossy channels with low delay tolerance such as roadside to vehicle safety, cellular, satellite networks and internet television (TV). In wireless networks, packet loss due to many phenomena related to the propagation environment and mobility are seen as packet erasure at higher communication layers and thus can be used by network coding to expedite the recovery process \cite{ref18}. In this paper, we are interested in a class of applications where all the receivers should receive all the needed packets with a predefined delay tolerance regardless of the packet order. Exceeded that delay, the frame is no longer needed.

An important NC subclass for the aforementioned application is the Instantly Decodable Network Coding (IDNC) \cite{ref2,ref3,ref4,ref5,ref6,ref8,ref13,ref14,ref17,ref18,arg1,refsameh,refahmed,refjournal,xiao1,xiao2}. This subclass can be implemented by using simple binary XOR to encode and decode packets and thus allows a fast encoding and decoding, eliminating the need for computationally expensive matrix inversions. Moreover, no buffer is needed at the receivers to store non-instantly decodable packets. These proprieties allow the design of cost efficient receivers and thus the proliferation of wireless networks. 

Due to the erasure nature of the links in a wireless network configuration that affects the delivery of meaningful data, the receivers are no longer able to synchronously decode the frame. Therefore a better use of the channel and network does not mean an effective better throughput at higher communication layers \cite{1208721}. Fundamental research has been conducted to better understand the delay aspects in IDNC and more generally in NC. These studies can mainly be divided into two groups where the delay is considered as the:
\begin{itemize}
\item Completion time: which is the overall transmission time.
\item Decoding delay: which is the individual delay when delivered an useless packet at its reception moment.
\end{itemize}

The completion time experienced when delivered a frame is composed of a fixed delay (the initial transmission phase of the frame) and a variable delay (the recovery phase). This definition of delay depends on the channel condition. If the channel condition is harsh (i.e. high erasure probability), the completion time will increase regardless if the receivers effectively received the packet or if it was erased. The completion time performance of IDNC was studied in \cite{ref4} in a case of perfect feedback and in \cite{ref13,refsameh,ref14,ref17} for limited feedback. The decoding delay offers a definition more independent of the channel conditions since no delay is taken into account if the intended packet is erased and only delays due to the chosen encoded packet are taken into account. The decoding delay performance of IDNC was studied for perfect feedback and memory-less channels (MEC) in \cite{ref2,ref8} then extended to persistent erasure channels (PEC) in \cite{ref3}. In \cite{refahmed} the authors studied this performance in a lossy intermittent feedback for MECs and in \cite{refjournal} extended it to PECs.

In all the aforementioned works which considered the delay in IDNC as the decoding delay, the authors considered the decoding delay as the sum of all the individual decoding delays experienced by all the receivers. This definition of delay do not permit to have equitable distribution of the delays between the different receivers since only the sum of all the individual delays counts. For application of our interest in this paper, the receivers have strict delay tolerance after which the frame is no longer needed such as streaming TV and cellular. Consequently, we introduce the following definition of delay in IDNC:
\begin{definition}
The delay experienced when sending a frame is the maximum decoding delay experienced by all the receivers.
\end{definition}

In this paper, we address the following question: What is the best policy to minimize the maximum decoding delay and how to extend IDNC algorithm to operate under these circumstances? In order to answer the former question, we first identify the maximum decoding delay increment probabilities. We then employ these expressions to formulate the minimize decoding delay as a maximum weight clique problem in the IDNC graph. We subsequently design a greedy algorithm to perform packet selection since the problem is known to be NP hard \cite{arg1}. Finally, we compare through extensive simulations the different delay aspects when using the different policies.

The rest of this paper is organized as follows. \sref{sec:sys} introduces the system model and parameters. In \sref{sec:formulation}, we compute the maximum decoding delay increment and introduce our problem formulation. Our proposed greedy algorithm is illustrated in \sref{sec:algo} before presenting the simulation results in \sref{sec:results}. Finally, we conclude this paper in \sref{sec:conclusion}.

\section{System Model and Parameters} \label{sec:sys}

Let a wireless sender that wishes to deliver a frame $\mathcal{N}$ of $N$ source packets to a set $\mathcal{M}$ of $M$ receivers. The receivers are interested in getting all packets of $\mathcal{N}$ within a fixed decoding delay $T$. After this delay the frame is no longer needed as in the multimedia streaming. The sender begins by transmitting the $N$ packets of the frame uncoded in an \emph{initial phase}. In \cite{refsameh}, it was shown that sending packets uncoded achieves a lower delay in the \emph{initial phase} than combination some of them. Each receiver and for each successfully received packet listens, an acknowledgment is transmitted to the sender. Whereas the packet transmissions are subject to loss, the feedback transmissions are assumed to be perfect. Let $p_{i}$, $i\in\mathcal{M}$, be the probability to loss a packet at receiver $i$ (i.e. packet erasure probability), assumed to be constant during the transmission.

At the end of the \emph{initial phase}, the packets of each receiver $i$ can be in one of the following sets:
\begin{itemize}
\item The \emph{Has} set $\mathcal{H}_i$: the set of packets correctly transmitted to receiver $i$.
\item The \emph{Wants} set $\mathcal{W}_i$: the set of packets erased at receiver $i$ and need to be resend. We have $\mathcal{N} =  \mathcal{W}_i \cup \mathcal{H}_i$.
\end{itemize}

These information are stored at the sender in a \emph{feedback matrix} $\mathbf{F} = [f_{ij}],~ \forall~ i \in \mathcal{M},~ \forall~j \in\mathcal{N}$ such that:
\begin{align}
f_{ij} =
\begin{cases}
0 \hspace{0.9 cm}& \text{if } j \in \mathcal{H}_i \\
1 \hspace{0.9 cm}& \text{if } j \in \mathcal{W}_i.
\end{cases}
\end{align}

To complete the transmission of the erased packets in the \emph{initial phase}, the sender applies binary XOR-network coded combinations of the source packets using the feedback matrix. In this phase, each receiver that received and decoded a packet sends an acknowledgement which is used by the sender to update the feedback matrix and the Has/Wants sets of that receiver. This process is repeated until all receivers feedback that they obtained all packets of the frame. In this \emph{recovery phase}, the packet combination for each receiver $i$ can be either:
\begin{itemize}
\item \emph{Non-innovative} if all the packets encoded in it were successfully received previously.
\item \emph{Instantly Decodable} if it contains a \emph{single} source packet from $\mathcal{W}_i$.
\item \emph{Non-Instantly Decodable} if it contains \emph{several} source packets from $\mathcal{W}_i$.
\end{itemize}

Therefore the decoding delay of receiver $i$, with non empty Wants set, increases if he successfully receives an non-innovative of non-instantly decodable packet.

\section{Minimum Decoding Delay Formulation} \label{sec:formulation}

\subsection{Maximum Decoding Delay Increment}

Let $d_i(\kappa,t)$ be the decoding delay increase for receiver $i$, at time $t$, after the transmission $\kappa$. Define the targeted receivers by a transmission as the receivers that can instantly decode a packet from that transmission. According to the analysis done in \cite{ref2}, the probability of the decoding delay increase for receiver $i$ with non-empty Wants set is
\begin{align}
\mathds{P}\left(d_i(\kappa,t) = 0 \right) =
\begin{cases}
1 \hspace{0.4 cm}& \text{if }i \text{ targeted by }\kappa \\
p_i \hspace{0.4 cm}& \text{if }i \text{ is not targeted by }\kappa.
\end{cases}
\end{align}

Define $D_i(n)$ as the total decoding delay experienced by receiver $i$ until the transmission at time $n$ (i.e. $D_i(n) = \sum\limits_{t=1}^n d_i(\kappa,t)$). It is clear that $D_i(n-1) \leq D_i(n),~\forall~n > 1$. Let $\mathds{X}(t)$ be the event that the maximum decoding delay increases at time $t$. The mathematical definition of the probability that this event occurs can be expressed as
\begin{align}
\mathds{P}(\mathds{X}(t)) &= \mathds{P}\left( \underset{i \in \mathcal{M}}{\text{max }}D_i(t-1) < \underset{i \in \mathcal{M}}{\text{max }} D_i(t) \right) \nonumber \\
&= 1 - \mathds{P}\left( \underset{i \in \mathcal{M}}{\text{max }}D_i(t-1) = \underset{i \in \mathcal{M}}{\text{max }} D_i(t) \right).
\end{align}

Define $\mathcal{L}(t)$ as the set of receivers having the maximal decoding delay at time $t$ (i.e. $i \in \mathcal{L}(t) \Leftrightarrow D_i(t) = \underset{j \in \mathcal{M}}{\text{max }}D_j(t)$). From the previous definitions, we have:
\begin{align}
&\mathds{P}\left( \underset{i \in \mathcal{M}}{\text{max }}D_i(t-1) = \underset{i \in \mathcal{M}}{\text{max }} D_i(t) \right) \\
& \qquad = \mathds{P} \left(d_i(\kappa,t) = 0 , \forall~i \in \mathcal{L}(t)\right)  \nonumber \\
&  \qquad = \prod_{i \in \mathcal{L}(t)} \mathds{P} (d_i(\kappa,t) = 0).\nonumber 
\end{align}

Let $\tau(\kappa)$ be the set of targeted receivers in the transmission $\kappa$ and $M_w$ the set of receivers having non-empty Wants set. The probability of event $\mathds{X}(t)$ to occur can be expressed as follows:
\begin{align}
\mathds{P}(\mathds{X}(t)) &= 1 - \prod_{i \in \mathcal{L}(t)} \mathds{P} (d_i(\kappa,t) = 0) \nonumber \\
&= 1 -  \prod_{i \in (\mathcal{L}(t) \cap M_w) \setminus \tau(\kappa)} p_i.
\end{align}

\subsection{Problem Formulation}

To represent all the possible combinations of source packets, we use the IDNC graph $\mathcal{G} (\mathcal{V},\mathcal{E})$ introduced in \cite{ref2}. This graph is constructed by generating a vertex $v_{ij} \in \mathcal{V}$ for each packet $j \in \mathcal{W}_i,~ \forall~ i \in \mathcal{M}$ and then connecting two vertices if the packet combination is instantly decodable for both receivers represented by these vertices. According to the analysis done in \cite{arg1}, the set of all packet combinations in IDNC is represented by all maximal cliques in $\mathcal{G}$. The sender applies binary XOR to all the packets identified by the vertices of a selected maximal clique in $\mathcal{G}$ to generate the coded packet to be send. The targeted receivers by this transmission are those identified by the vertices of the selected maximal clique.

Given this IDNC graph formulation and from the previous expressions, we can express the minimum decoding delay problem as a maximum weight clique problem in the IDNC graph, such that
\begin{align}
\kappa^{*}(t) &= \underset{\kappa(t) \in \mathcal{G}}{\text{argmin}}\left\{ \mathds{P}(\mathds{X}(t))\right\} \nonumber \\
&= \underset{\kappa(t) \in \mathcal{G}}{\text{argmin}} \left\{ 1 -  \prod_{i \in (\mathcal{L}(t) \cap M_w) \setminus \tau(\kappa)} p_i \right\}   \nonumber \\
&= \underset{\kappa(t) \in \mathcal{G}}{\text{argmax}} \left\{ \prod_{i \in (\mathcal{L}(t) \cap M_w) \setminus \tau(\kappa)} p_i \right\}  \\
&= \underset{\kappa(t) \in \mathcal{G}}{\text{argmin}} \left\{\prod_{i \in \mathcal{L}(t) \cap \tau(\kappa)} p_i \right\} \nonumber \\
&= \underset{\kappa(t) \in \mathcal{G}}{\text{argmax}} \sum_{i \in \mathcal{L}(t) \cap \tau(\kappa)} \text{ln}\left(\cfrac{1}{p_i}\right).\nonumber
\end{align}

In other words, the minimum decoding delay problem can be formulated as a maximum weight clique problem where the weight of each vertex $v_{ij}$ can be expressed as:
\begin{align}
w_{ij}^* = \text{log}\left(\cfrac{1}{p_i}\right)= -log(p_i).
\label{origin}
\end{align}

\section{Proposed Heuristic Algorithm} \label{sec:algo}

Finding the maximum weight clique in the IDNC graph is shown to be NP-hard in \cite{arg1}. In order to overcome this complexity, we introduce a simple greedy algorithm to perform the packet selection over the IDNC graph. This algorithm is an extended multilayer version of the algorithm proposed in \cite{ref2,ref3,refsameh,refahmed}.

Let $w_{ij}$ be the modified weights. These weights reflect a high original weight of the vertex and a high connection to vertices having high original weight. The mathematical formulation of these weights is given by:
\begin{align}
w_{ij} = (w_{ij}^* + 1) \times \sum_{v_{kl} \in \mathcal{V}_{ij}} w_{kl}^*,
\label{omegamax}
\end{align}
where $\mathcal{V}_{ij}$ is the set of vertices connected to $v_{ij}$.
\begin{algorithm}[t]
\begin{algorithmic}
\REQUIRE $\mathbf{F}$, $p_i \text{ and } D_{i},~\forall~ i\in\mathcal{M}$.
\STATE Initialize $\kappa^* =\varnothing$.
\STATE Construct $\mathcal{G}_1\left(\mathcal{V}_1,\mathcal{E}_1\right), \mathcal{G}_2\left(\mathcal{V}_2,\mathcal{E}_2\right), ..., \mathcal{G}_h\left(\mathcal{V}_h,\mathcal{E}_h\right)$.
\FOR{l=1 \TO h}
\STATE $\mathcal{G} \leftarrow \mathcal{G}_l$.
\FORALL{ $v \in \kappa^*$}
\STATE Sets $\mathcal{G} \leftarrow \mathfrak{R}(\mathcal{G},v)$.
\ENDFOR
\WHILE{$\mathcal{G} \neq \varnothing$}
\STATE Compute $w_{ij}^*$ and $w_{ij}$ using \eref{origin} and \eref{omegamax}.
\STATE Select $v^* =\underset{v_{ij}\in\mathcal{G}}{\text{argmax}} \left\{w_{ij}\right\}$.
\STATE Sets $\kappa^* \leftarrow \kappa^* \cup v^*$.
\STATE Sets $\mathcal{G} \leftarrow \mathfrak{R}(\mathcal{G},v^*)$.
\ENDWHILE
\ENDFOR
\end{algorithmic}
\caption{Maximum Weight Vertex Search Algorithm}
\label{algo1}
\end{algorithm}
Let $\mathcal{L}_1(t),\mathcal{L}_2(t),...\mathcal{L}_h(t)$ be the sets of vertices with $h\leq M$ such that:
\begin{itemize}
\item $(v_{ij}, v_{kl}) \in \mathcal{L}_n \Leftrightarrow D_i(t) = D_k(t),~\forall~n \leq h$.
\item $v_{ij} \in \mathcal{L}_m$ and $v_{kl} \in \mathcal{L}_n \Leftrightarrow D_i(t) < D_k(t), ~\forall~n < m \leq h$.
\end{itemize}
From the definitions above, it is clear that $\mathcal{L}(t)= \mathcal{L}_1(t)$. We define $\mathcal{G}_i\left(\mathcal{V}_i,\mathcal{E}_i\right)$ as the graph associated to the set of receivers in $\mathcal{L}_i(t),~\forall~ i \leq h$ and let $\mathcal{G}\left(\mathcal{V},\mathcal{E}\right)$ be the global graph (i.e. $\mathcal{G}\left(\mathcal{V},\mathcal{E}\right)= \bigcup\limits_{i = 1 }^h \mathcal{G}_i\left(\mathcal{V}_i,\mathcal{E}_i\right)$). 

In order to minimize the maximum decoding delay experienced by the receivers, we apply the  maximum weight vertex search algorithm on the sub-graph $\mathcal{G}_1\left(\mathcal{V}_1,\mathcal{E}_1\right)$, containing the vertices having the maximum decoding delay so far, to obtain the maximal weight clique $\kappa^*$ in that graph. We, then, construct $\mathfrak{R}(\mathcal{G}_2,v), ~\forall~ v \in \kappa^*$, where $\mathfrak{R}(\mathcal{G},v_{ij})$ is the subgraph in $\mathcal{G}$ containing only the vertices connected to $v_{ij}$. This sub-graph represents the vertices that can eventually increase the maximum decoding delay after $2$ transmissions (in general after $Di(t)-Dk(t)+1$ where $i\in \mathcal{G}_1$ and $i\in \mathcal{G}_2$). The maximal weight clique in that graph is selected using the algorithm then added to the first one since they are combinable. This process is repeated for each layer $\mathcal{G}_i, i\leq h$ of the graph to find the selected maximal weight clique $\kappa^*$ to be served over the whole graph that guarantee the minimum expected increase in the maximum decoding delay. The entire algorithm structure is illustrated in Algorithm \ref{algo1}.

\section{Simulation Results} \label{sec:results}

In this section, we first present the simulation results comparing the different delays aspects achieved by the different policies to optimize each. We then present the performance of our policy to reduce the maximum decoding delay, against the sum decoding delay policy, to serve receivers having strict delay constraints.

In the first part, we compare, through extensive simulations the sum decoding delay (SDD) and the maximum decoding delay (MDD) experienced by the receivers while using the policy to reduce the SDD and our policy to reduce the MDD with the assumption that the receivers do not have delay constraints (i.e. $T=\infty$).

\begin{figure}[t]
\centering
  \includegraphics[width=1\linewidth,height=180pt]{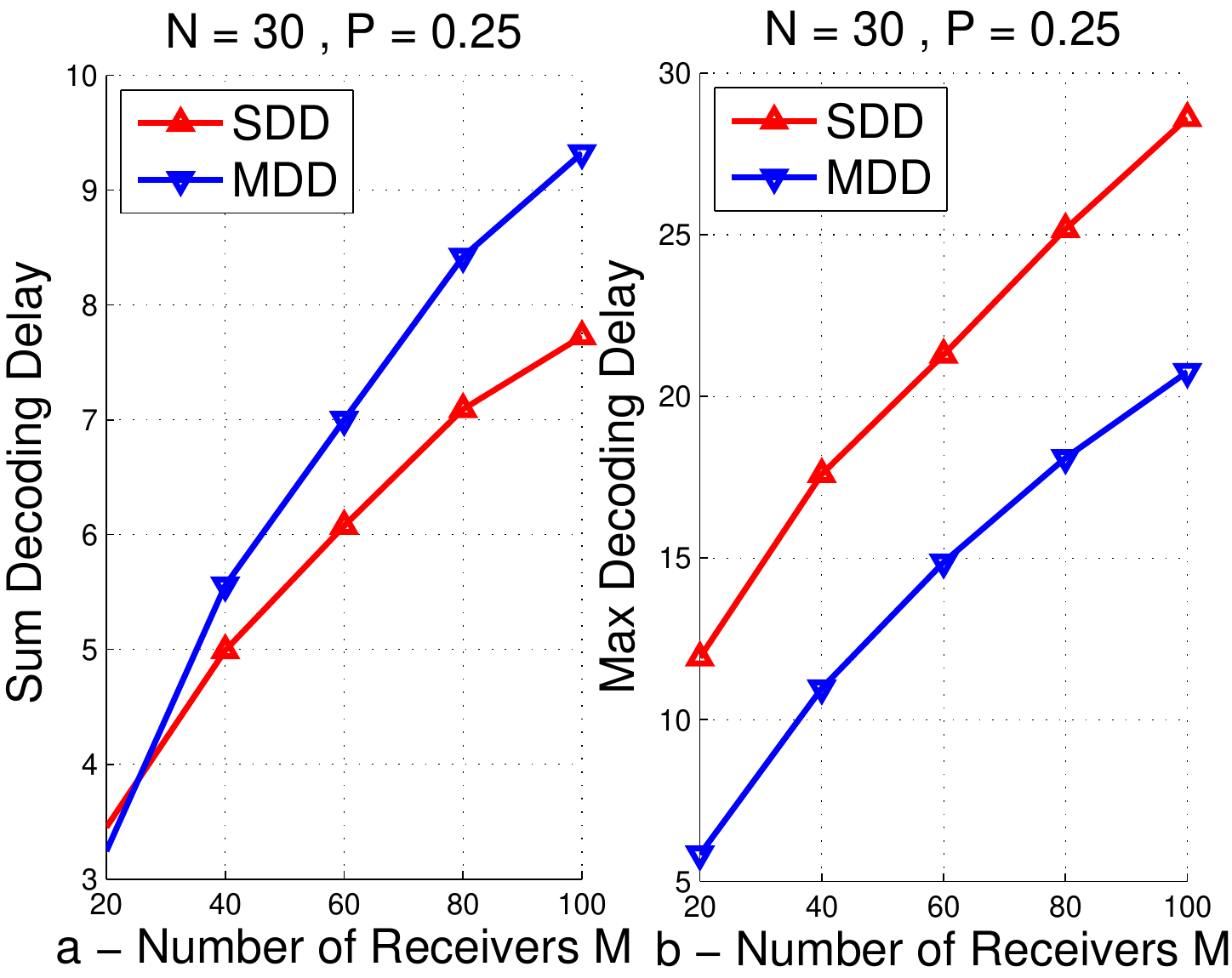}\\
  \caption{Mean delays for IDNC versus $M$ for a low erasure channel.}\label{fig:M}
\end{figure}

\begin{figure}[t]
\centering
  \includegraphics[width=1\linewidth,height=180pt]{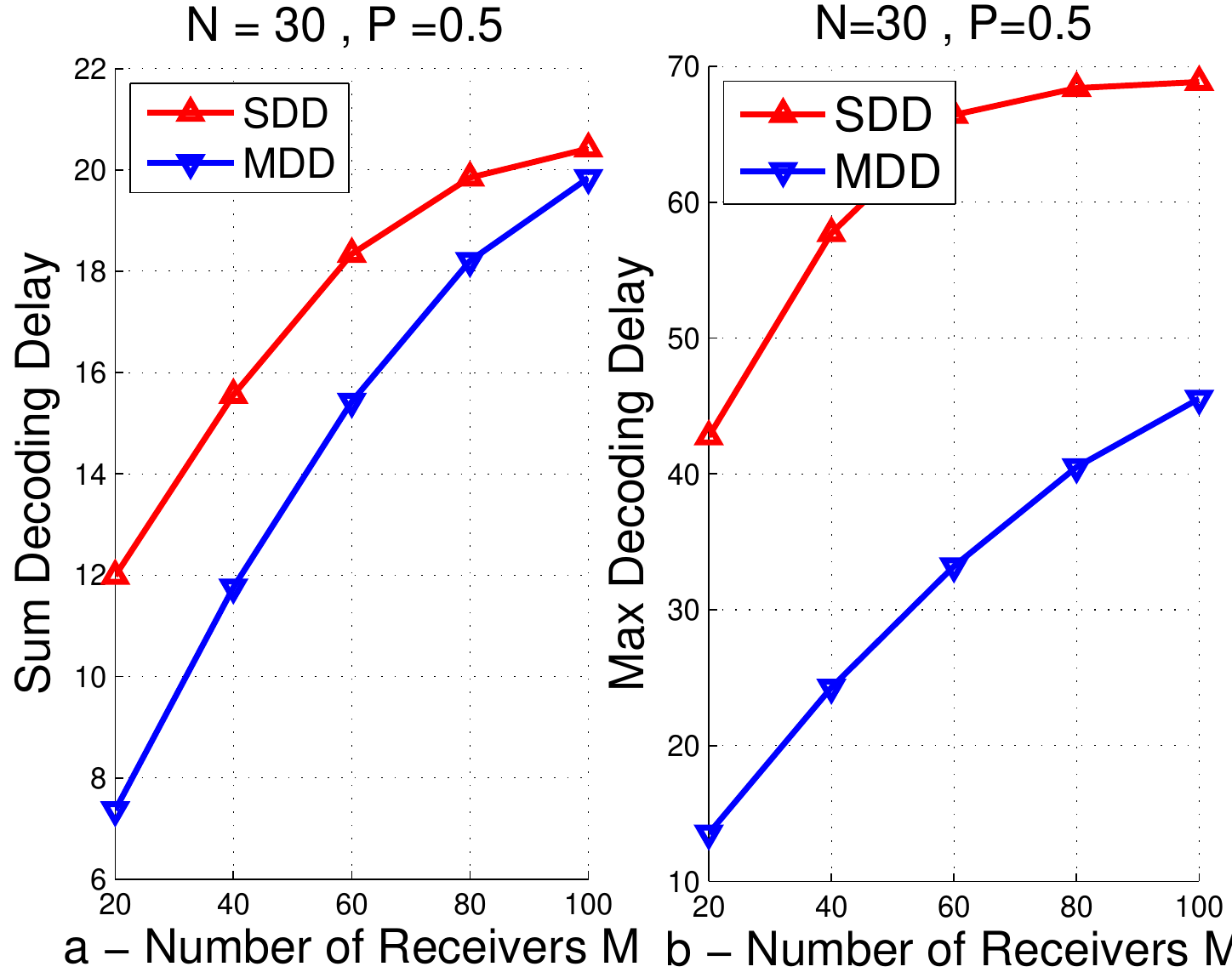}\\
  \caption{Mean delays for IDNC versus $M$ for a high erasure channel.}\label{fig:M2}
\end{figure}

\begin{figure}[t]
\centering
  \includegraphics[width=1\linewidth,height=180pt]{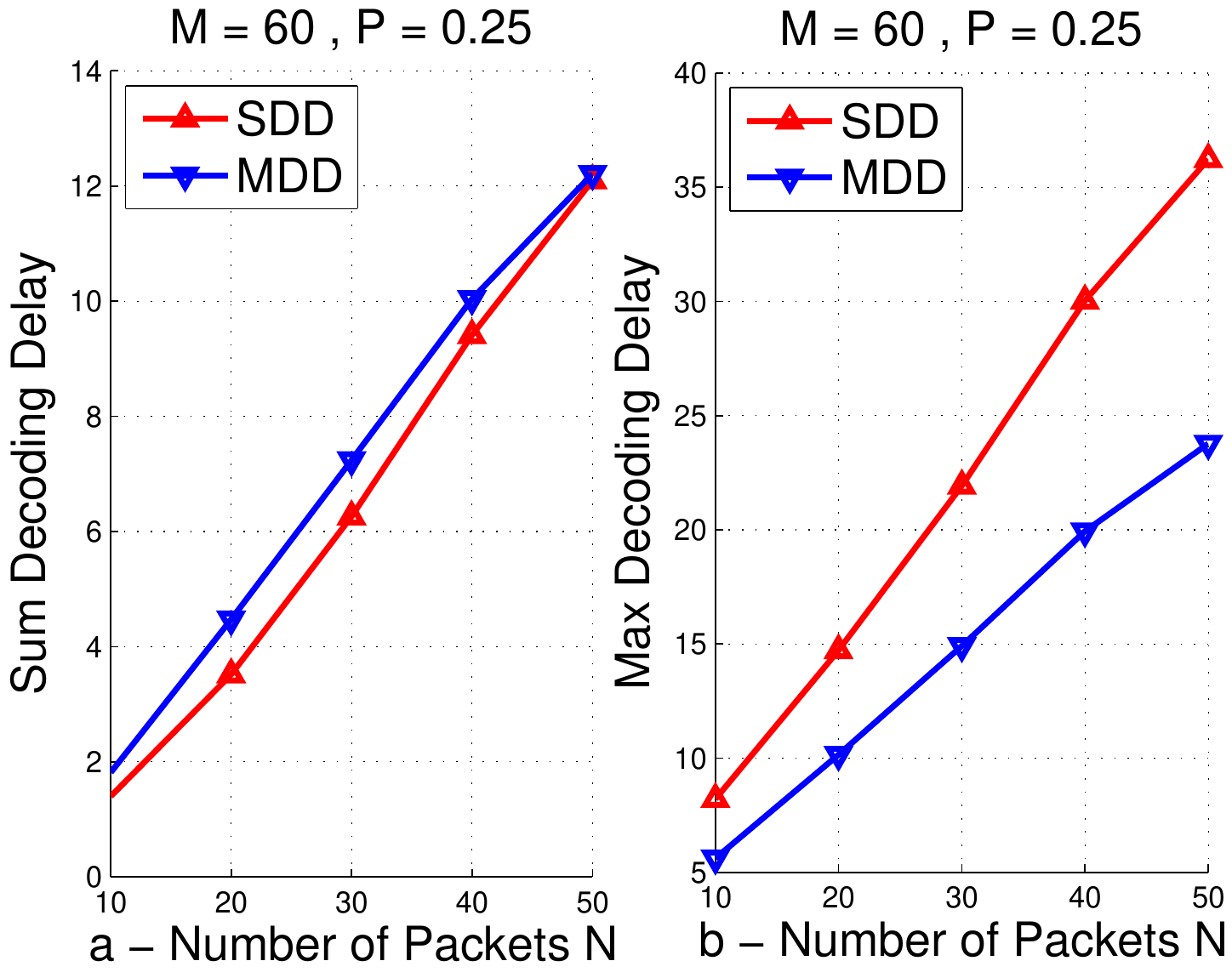}\\
  \caption{Mean delays for IDNC versus $N$ for a low erasure channel.}\label{fig:N}
\end{figure}

\begin{figure}[t]
\centering
  \includegraphics[width=1\linewidth,height=180pt]{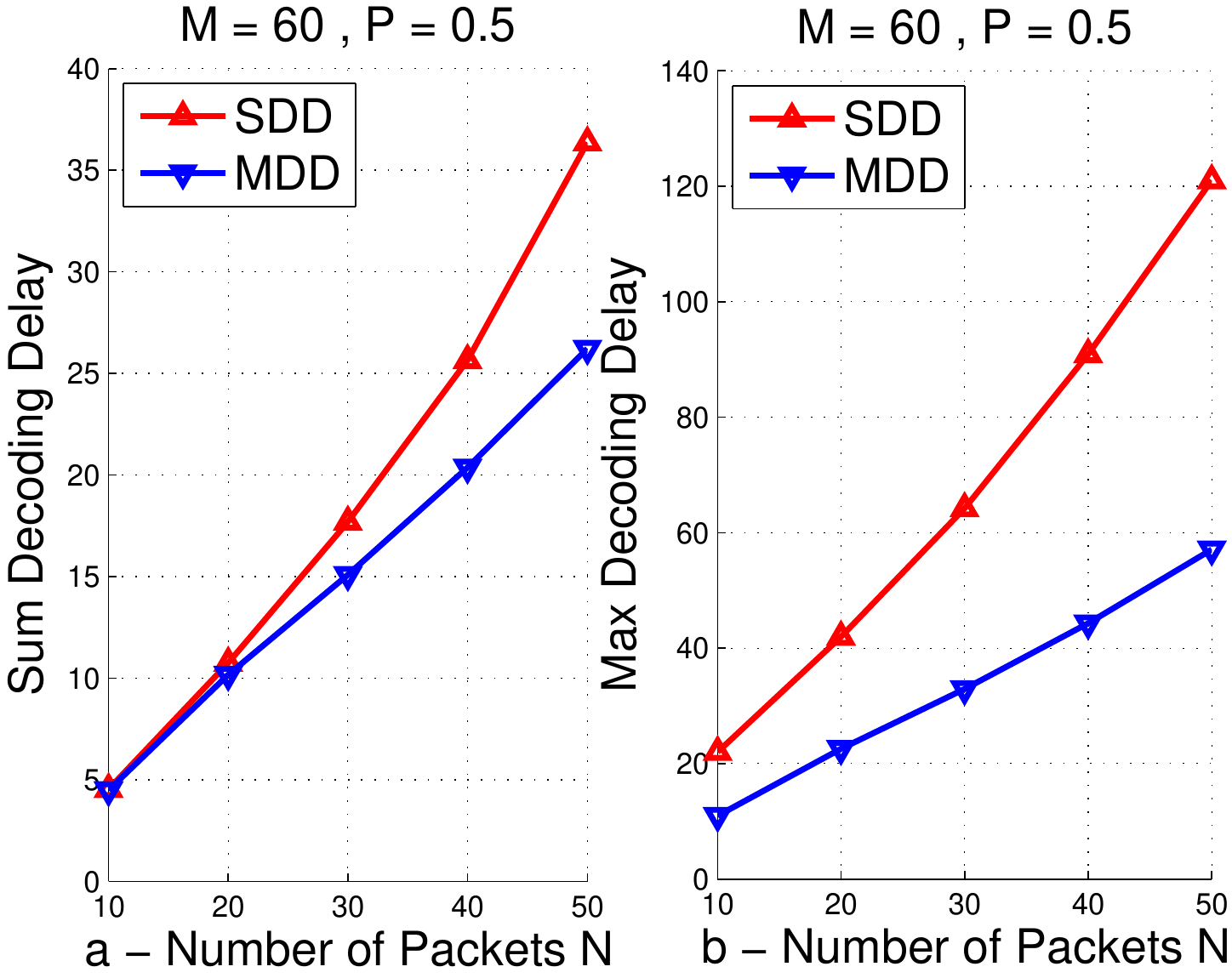}\\
  \caption{Mean delays for IDNC versus $N$ for a high erasure channel.}\label{fig:N2}
\end{figure}

\begin{figure}[t]
\centering
  \includegraphics[width=1\linewidth,height=180pt]{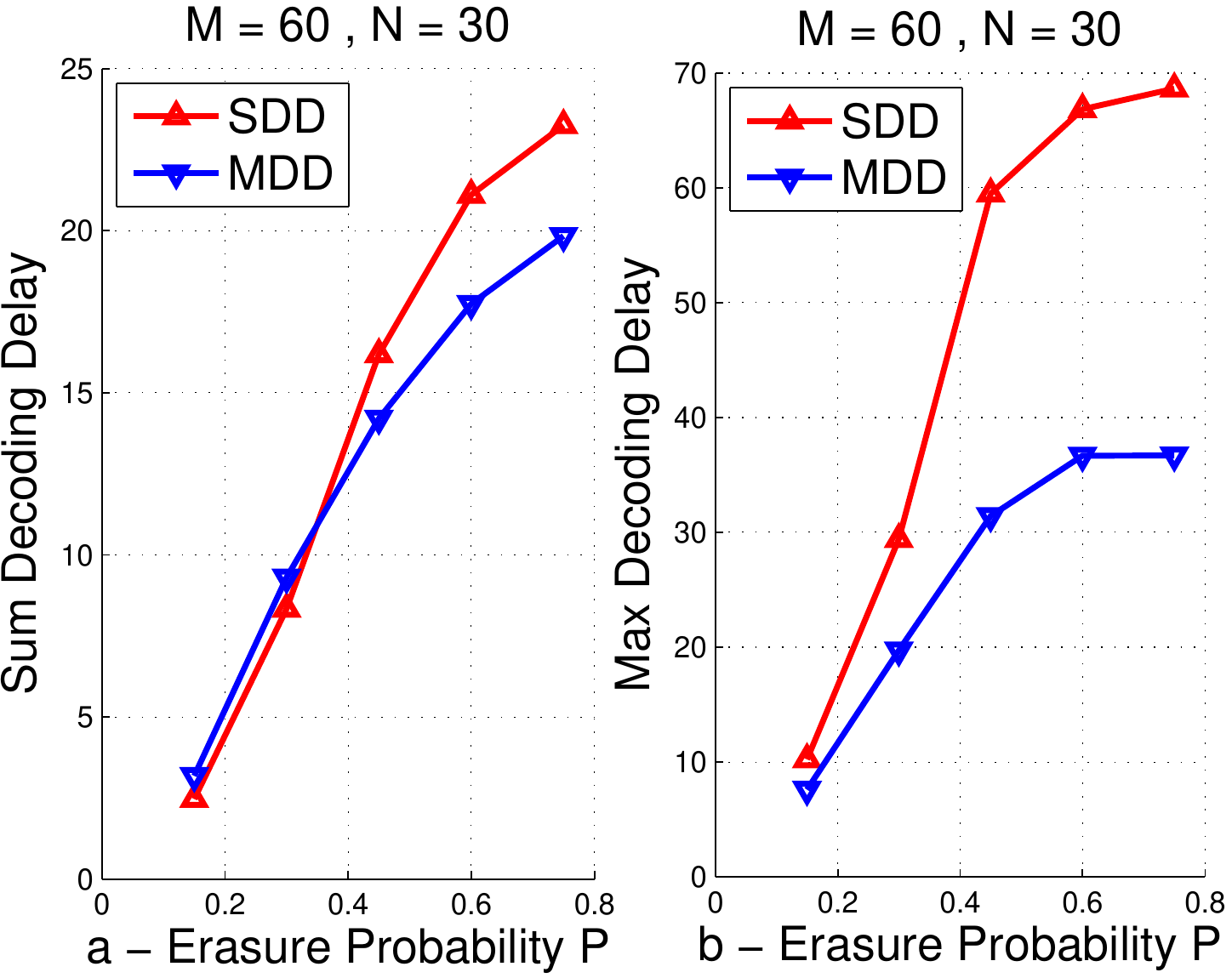}\\
  \caption{Mean delays for IDNC versus packet erasure probability $P$.}\label{fig:P}
\end{figure}

\begin{figure}[t]
\centering
  \includegraphics[width=1\linewidth,height=180pt]{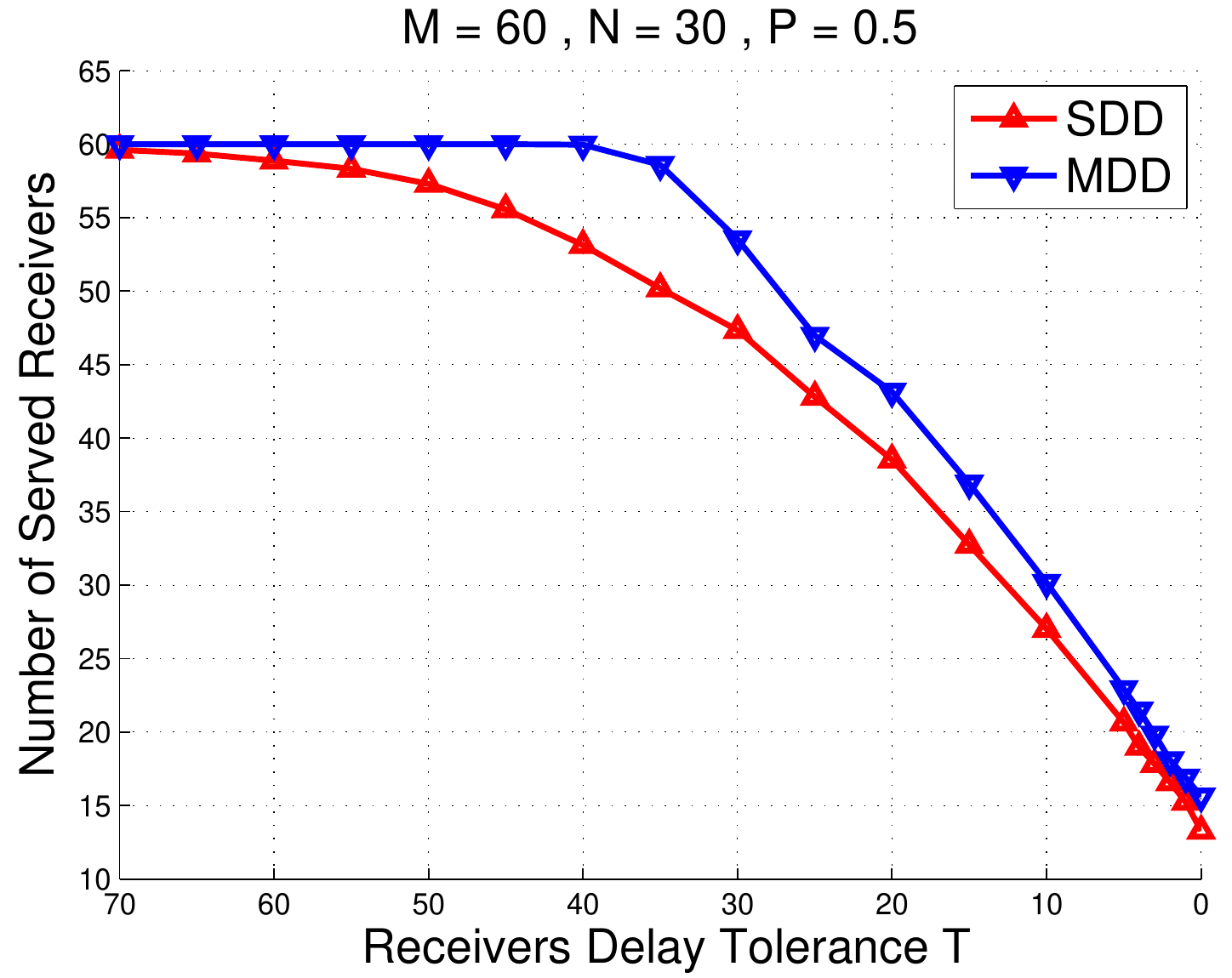}\\
  \caption{Mean number of served receivers versus the delay contraint $T$.}\label{fig:MAXREC}
\end{figure}

In the second part, we compare the number of receivers successfully served using the different policies while increasing the delay constraints of the different receivers. In all the simulations, the different delays are computed by frame then averaged over a large number of iterations. We assume that the packet erasure probability of all the receivers change from frame to frame while the average packet erasure probability $P$ remain constant.

\fref{fig:M}.a depicts the comparison of the mean sum decoding delay achieved by the policy to reduce the sum decoding delay (SDD) and the one to reduce the maximum decoding delay (MDD) against $M$ for $N=60$ and $P=0.25$. \fref{fig:M}.b illustrates the comparison of the max decoding delay for the same inputs. \fref{fig:M2} shows the same comparison than \fref{fig:M} but for high erasure channel ($P=0.5$). \fref{fig:N} and \fref{fig:N2} depicts the comparison of the aforementioned delay aspects against $N$ for $M=60$ and $P=0.25$ and $P=0.5$ receptively and \fref{fig:P} illustrates this comparison against the erasure probability $P$ for $M=60$ and $N=30$. \fref{fig:MAXREC} shows the number of successfully served receivers achieved by the SDD and the MDD policies against the delay constraint $T$ for $M=60,N=30$ and $P=0.5$.

From all the figures, we can clearly see that our proposed definition of delay gives the best agreement among the decoding delay definitions in IDNC. The maximum decoding delay policy offers, in average, the minimum sum of all the delay aspects in all situations.

\fref{fig:M}.a and \fref{fig:N}.a depicts the sum decoding delay when applying the maximum decoding delay policy and the sum decoding delay policy against $M$ and $N$ for a low packet erasure probability. We see that the performance of MDD and SDD are very close. Whereas in \fref{fig:M}.b and \fref{fig:N}.b  where the maximum decoding delay is computed for the same inputs, the performance of MDD is much better than SDD one.

As the channel conditions become harsher (high packet erasure probability), our policy to reduce the maximum decoding delay minimize the sum decoding delay better than the SDD. We can see from \fref{fig:M2}.a, \fref{fig:M2}.b, \fref{fig:N2}.a and \fref{fig:N2}.b that MDD outperforms SDD in minimizing both the sum decoding delay and the maximum decoding delay. \fref{fig:P}.a shows that for $P>0.35$, MDD becomes the best policy to effectively reduce all the decoding delay aspects in IDNC. This can be explained by the light of the SDD policy characteristics. In the SDD policy, the heuristic algorithm to perform the maximum weight clique problem selects the vertices with the highest reception probability. For a low erasure probability, the difference between the packet erasure probabilities of the different receivers is low and the maximum weight clique can be approximated by this heuristic. However, for a higher erasure probability, the maximum weight clique can be different since the difference between the erasure of receivers is significant. As a consequence the selection is no longer effective. In contrast, our policy always performs an effective packet selection since it relies not only on the reception probability but also on the decoding delay experienced so far by each receiver which is different in all situations.

\fref{fig:MAXREC} illustrates the performance of our policy to serve receivers with strict delay constraint against the SDD policy. We clearly see that MDD does not experience degradation until a certain delay constraint unlike SDD. For example MDD serves all the receivers ($100\%$) until the delay constraint $T=40$ whereas SDD serves only $90\%$ with this delay constraint.

\section{Conclusion} \label{sec:conclusion}

In this paper, we first introduced the maximum decoding delay as an alternative definition of delay for generalized instantly decodable network coding. We studied the problem of minimizing the maximum broadcast decoding delay experienced by all the receivers of IDNC. The MDD policy offers a better dividing of the delay experienced by the different receivers, unlike the sum decoding delay. We derived the expressions for the probability distributions of the maximum decoding delay increments and used them to formulate the problem as a maximum weight clique problem in the IDNC graph. In order to solve the problem in linear time with the size of the graph, we designed a greedy multilayer algorithm to perform effective packet selection. Simulations results showed that the MDD policy not only offers a good compromise among all the aforementioned decoding delay aspects in all situations and outperforms the sum decoding delay policy to effectively minimize the sum decoding delay when the channel conditions become harsher but also improves significantly  the number of served receivers when they are subjected to strict delay constraints.

\appendices

\bibliographystyle{IEEEtran}
\bibliography{references}

\end{document}